\documentclass[conference]{IEEEtran}
\IEEEoverridecommandlockouts

\usepackage{xurl}
\usepackage{hyperref}
\hypersetup{
    colorlinks=true,
    urlcolor=cyan,
    linkcolor=black,
    filecolor=black, 
    citecolor=black,
    pdfpagemode=FullScreen,
    }

\usepackage{colortbl}
\usepackage[compress]{cite}
\usepackage{amsmath,amssymb,amsfonts}
\usepackage{algorithmic}
\usepackage{graphicx}

\usepackage{subcaption}
\usepackage{multirow}
\usepackage{booktabs}
\usepackage{color,soul}
\usepackage{stix}
\usepackage[table]{xcolor}

\usepackage{tcolorbox}
\usepackage{dblfloatfix}
\usepackage{booktabs}
\usepackage[inline]{enumitem}
\newtcolorbox{boxH}{
    colback = white!90!gray, 
    colframe = black, 
    boxrule = 0pt, 
    leftrule = 3pt
}

\newtcolorbox{boxH2}{
    colback = green!10, 
    colframe = black, 
    boxrule = 0pt, 
    leftrule = 3pt,
    top = 1pt,
    bottom = 1pt
}
\newtcolorbox{boxH3}{
    colback = green!10, 
    colframe = black, 
    boxrule = 0pt, 
    leftrule = 3pt,
    left = 1pt,
    right = 1pt,
    top = 1pt,
    bottom = 1pt
}

\def\BibTeX{{\rm B\kern-.05em{\sc i\kern-.025em b}\kern-.08em
    T\kern-.1667em\lower.7ex\hbox{E}\kern-.125emX}}
\begin{document}

\title{TestMiner: Software Testing Analysis for\\GitHub Repositories}

 \author{
 \IEEEauthorblockN{Andre Hora}
 \IEEEauthorblockA{
\textit{Department of Computer Science, UFMG}\\
 Belo Horizonte, Brazil \\
 andrehora@dcc.ufmg.br}
  \and
  \IEEEauthorblockN{José Miguel Rojas}
 \IEEEauthorblockA{
\textit{University of Sheffield}\\
 Sheffield, UK \\
 j.rojas@sheffield.ac.uk}

 \and
  \IEEEauthorblockN{Romain Robbes}
 \IEEEauthorblockA{
\textit{University of Bordeaux, CNRS}\\
 Bordeaux, France \\
 romain.robbes@labri.fr}
 }


\maketitle

\begin{abstract}

Software systems have unique testing characteristics.
Some projects can emphasize unit tests, while others may focus on end-to-end testing.
Test organization may vary across ecosystems: in languages like Python and Java, tests are typically placed in dedicated folders, whereas Go and Rust projects commonly co-locate tests with source code.
These distinctions make it harder to understand how a project approaches testing.
In this paper, we present TestMiner, a tool for exploring software testing in GitHub repositories.
TestMiner provides an overview of a project’s testing practices, including test statistics, test location, test metrics across releases, and dependencies related to testing.
We used TestMiner in an undergraduate Software Testing course, where 50 students explored the testing practices of real-world GitHub repositories.
Overall, students expressed positive feedback regarding TestMiner.
They were able to critically explore a variety of testing practices, including test organization, test evolution, test fixtures, mocking, and edge-case testing.
TestMiner is available at: \url{https://andrehora.github.io/testminer}.
Screencast: \url{https://youtu.be/w1sBgLTq-7Y}.
\end{abstract}


\section{Introduction}

Software testing is a key activity in modern software development, and a good test suite is fundamental to ensuring software quality and sustainable evolution~\cite{beck2003test, hora2024test}.

Software systems have unique testing characteristics.
While some projects lean heavily on unit tests, others may focus more on end-to-end testing.
Projects may also adopt specific techniques, such as mocking, or integrate testing into CI/CD pipelines.
Test organization also varies across ecosystems: in languages like Python and Java, tests are typically placed in dedicated folders, whereas Go and Rust projects commonly co-locate tests with source code.
Additionally, projects may rely on multiple dependencies to support testing.

In practice, these distinctions make it harder to understand how a project approaches testing.
Developers onboarding to a project must explore the repository manually to understand its testing practices, a costly process that scales poorly as codebases grow.
Students seeking to understand real-world software testing face the same obstacle: testing practices are present in the repository, but they are not visible without an in-depth analysis. This gap motivates our work.

\begin{figure}[t]
    \centering
    \includegraphics[width=0.45\textwidth]{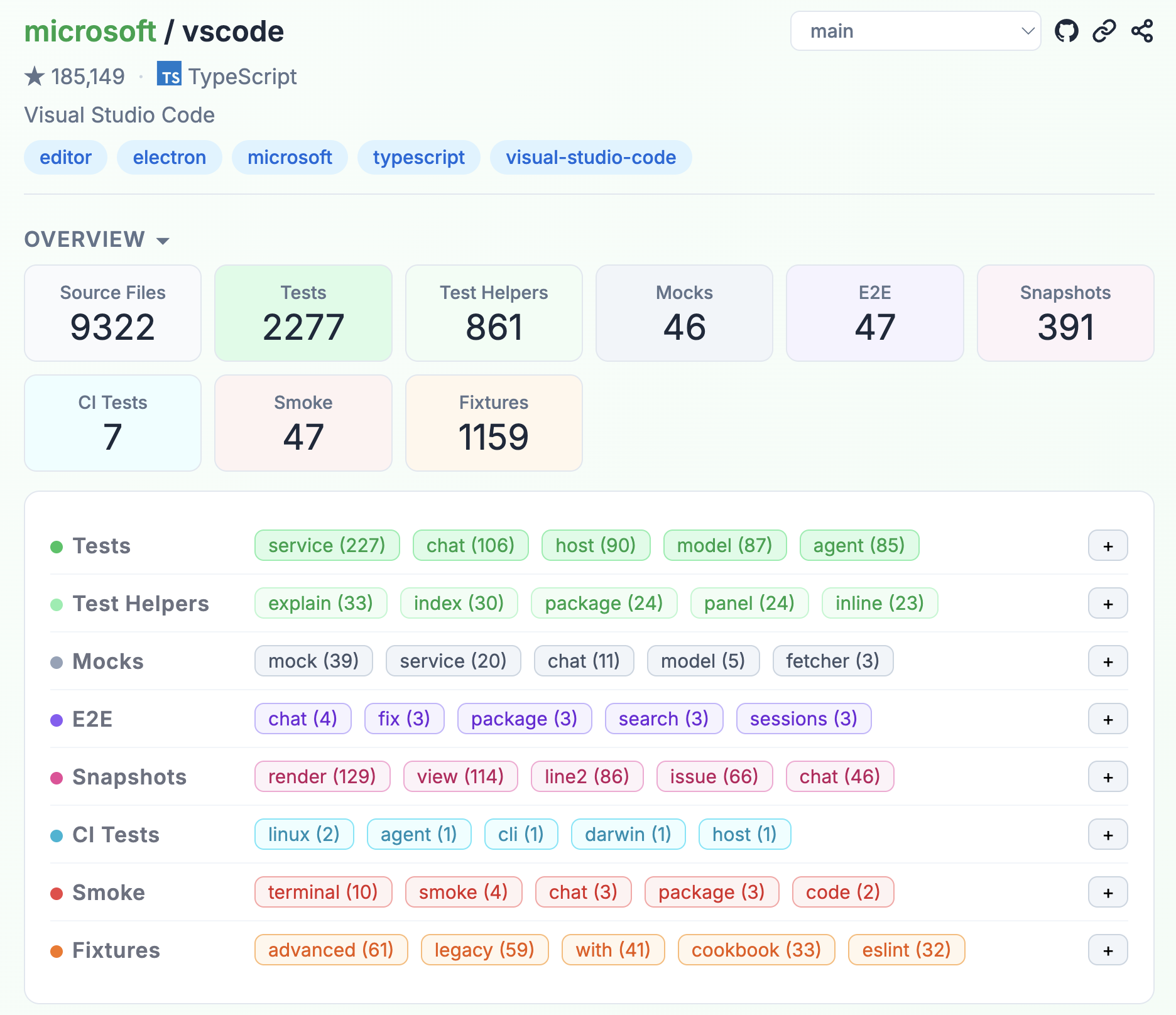}
    \caption{TestMiner overview for microsoft/vscode.}
    \label{fig:overview}
\end{figure}

This paper presents TestMiner, a tool for exploring software testing practices in GitHub repositories.
TestMiner provides an overview of a project’s testing practices, including test statistics, test location within the repository, test metrics across releases, and dependencies related to testing.
TestMiner is a web application and relies on GitHub and jsDelivr APIs to retrieve repository metadata and file-level information.
Users can enter any public GitHub repository, owner, or topic, and TestMiner processes the input to generate multiple testing reports, as the one presented in Figure~\ref{fig:overview} for microsoft/vscode.


We used TestMiner in an undergraduate Software Testing course, where 50 students explored the testing practices of real-world GitHub repositories~\cite{exploring-software-testing}.
Overall, students expressed positive feedback regarding TestMiner to understand software testing.
They were able to critically explore a variety of testing practices, including test organization, test evolution, test fixtures, mocking, and edge-case testing.

\begin{boxH}
\textbf{TestMiner}: \url{https://andrehora.github.io/testminer}

\textbf{Screencast}: \url{https://youtu.be/w1sBgLTq-7Y}
\end{boxH}


\begin{figure*}[t]
    \centering
    \includegraphics[width=0.9\textwidth]{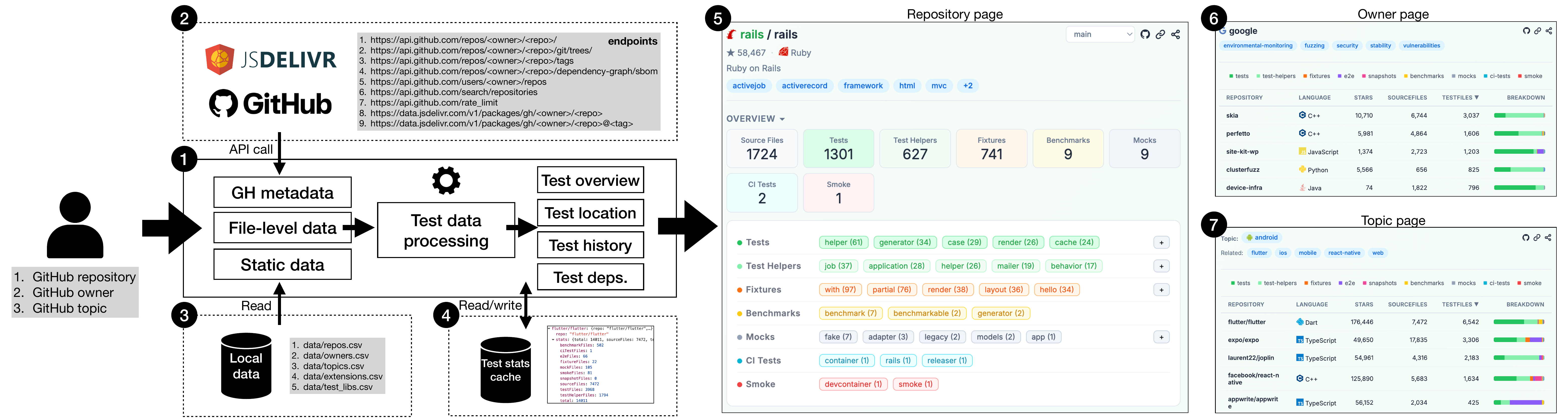}
    \caption{Overview of TestMiner.}
    \label{fig:testminer}
\end{figure*}

\noindent\textbf{Novelty.}
General repository mining frameworks~\cite{spadini2018pydriller, gitevo, Dabic:msr2021data, gitpython, isomorphic-git, jgit} provide no test-specific functionality out of the box, existing testing platforms, such as Codecov~\cite{CodeCov} and SonarQube~\cite{SonarQube}, require CI instrumentation, and test smell detectors, such as TSDetect~\cite{tsDetect} and PyNose~\cite{wang2021pynose}, provide per-test analyses only and require local setups.
The novelty of TestMiner lies in its ability to combine multiple testing analyses in a zero-installation, browser-based interface, allowing users to analyze any public GitHub repository without cloning or pipeline configuration. Our classroom evaluation provides initial evidence that this design lowers the effort required to understand how real-world open-source systems are tested.

\section{TestMiner}

\subsection{Overview}

TestMiner is a web application that runs in the browser--
Figure~\ref{fig:testminer} depicts a design overview.
(1) Users can enter a GitHub repository URL, GitHub owner, or GitHub topic. The input is processed by TestMiner, which generates multiple testing reports.
(2) TestMiner relies on GitHub and jsDelivr APIs to retrieve repository metadata and file-level information.
(3) Other data, such as autocomplete suggestions, comes from local sources.
(4) To avoid frequent requests to the external APIs, TestMiner caches selected testing statistics.
Depending on the user input, TestMiner generates three types of report pages: (5) repository, (6) owner, and (7) topic pages.

\subsection{Main Features}

\subsubsection{Repository View}
It provides a detailed testing analysis for a single repository and includes four sections: test overview, test location, test history, and test dependencies.

The \emph{Test Overview} section presents testing statistics for the analyzed repository.
To compute them, TestMiner retrieves the repository's file list via the jsDelivr API and classifies each file into one of ten categories: tests, test-helpers, fixtures, e2e, snapshots, benchmarks, mocks, ci-tests, smoke, and source code.
TestMiner classifies each file by applying a series of filename- and path-based rules.
Figures~\ref{fig:overview} and~\ref{fig:testminer}(5) show examples of the \emph{Test Overview} section TestMiner generated for microsoft/vscode\footnote{\url{https://andrehora.github.io/testminer/\#microsoft/vscode}} and rails/rails,\footnote{\url{https://andrehora.github.io/testminer/\#rails/rails}} respectively.

\begin{figure}[t]
    \centering
    \includegraphics[width=0.43\textwidth]{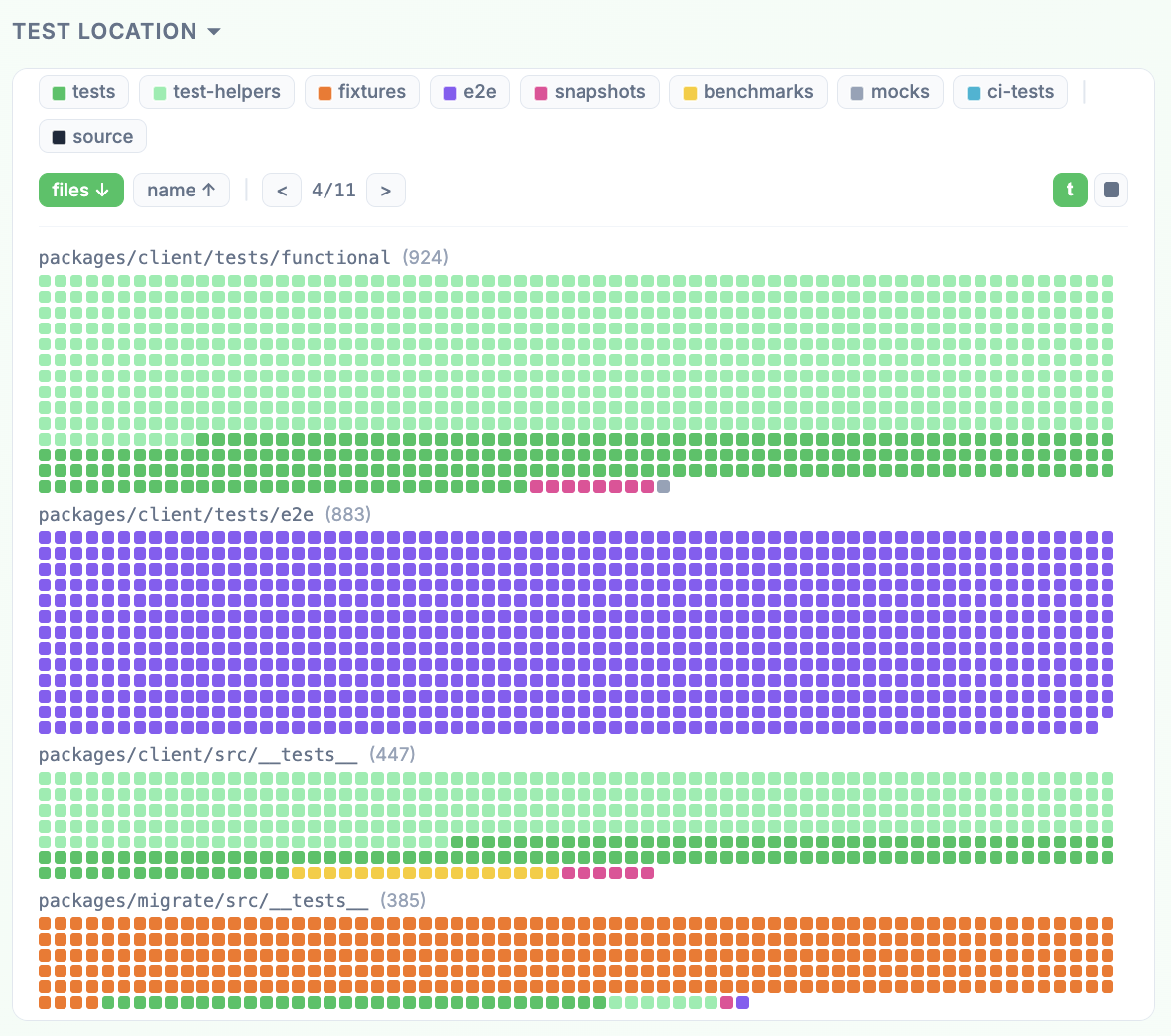}
    \caption{Test location for prisma/prisma.}
    \label{fig:test-location}
\end{figure}

The \emph{Test Location} section provides a file tree visualization showing where tests are located within the repository.
This visualization is inspired by Distribution Map, a generic technique to reason about the result of software analysis and to help to understand how a given phenomenon is distributed across a software system~\cite{ducasse2006distribution, silva2015developers, hora2012bug}.
It enables users to easily identify test locations, even in complex projects with deeply nested folder structures.
Users can also filter by test category, sort results, and adjust the depth of the folder structure displayed in the visualization.
Figure~\ref{fig:test-location} presents the \emph{Test Location} section for the project prisma/prisma.\footnote{\url{https://andrehora.github.io/testminer/\#prisma/prisma}}

\begin{figure}[t]
    \centering
    \includegraphics[width=0.43\textwidth]{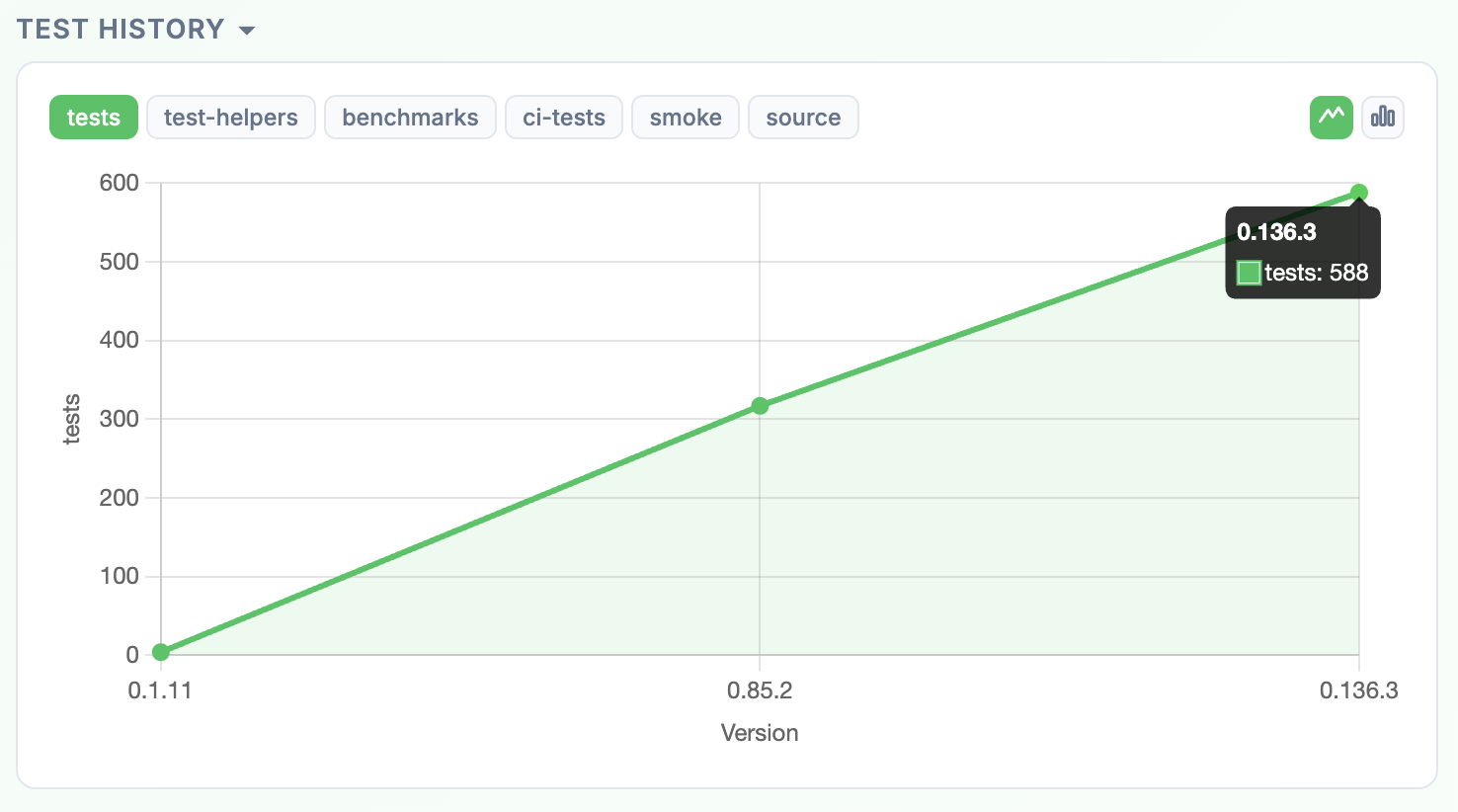}
    \caption{Test history for fastapi/fastapi.}
    \label{fig:test-hist}
\end{figure}

The \emph{Test History} section presents a chart showing testing metrics across three releases: the initial release, a middle release, and the most recent release.
It allows users to track the evolution of testing over time.
Figure~\ref{fig:test-hist} shows the \emph{Test History} section for project fastapi/fastapi, which increases from close to zero test files in the initial release to 300 in the middle release and nearly 600 in the most recent release.\footnote{\url{https://andrehora.github.io/testminer/\#fastapi/fastapi}}

\begin{figure}[t]
    \centering
    \includegraphics[width=0.43\textwidth]{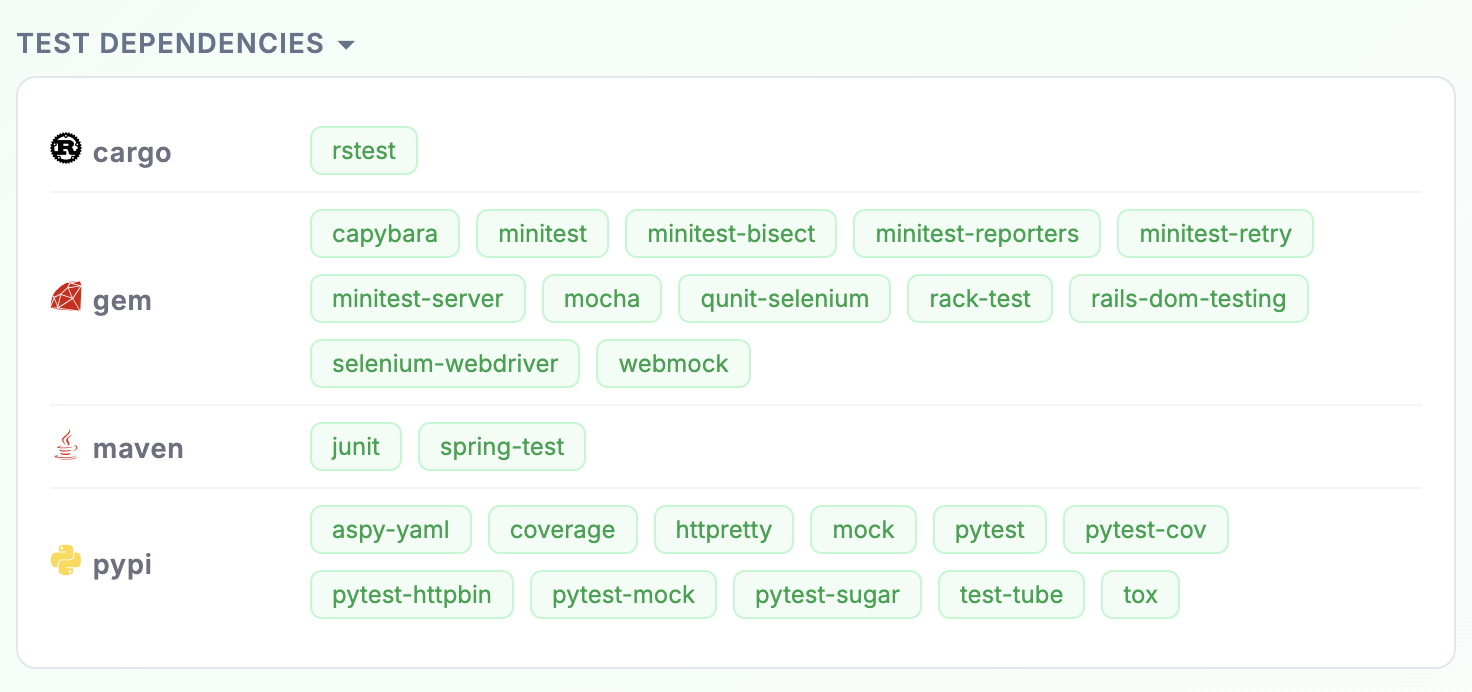}
    \caption{Test dependencies for github/linguist.}
    \label{fig:test-dep}
\end{figure}

Finally, the \emph{Test Dependencies} section details dependencies related to testing.
In this case, the dependencies are extracted from the GitHub Software Bill of Materials (SBOM), which lists the open-source and proprietary components that constitute a software product~\cite{nocera2025adoption, nocera2023software}.
GitHub's SBOM computation supports the analysis of dependencies in multiple ecosystems, such as Python, JavaScript, TypeScript, Java, Ruby, Go, Rust, and .NET.
Figure~\ref{fig:test-dep} shows the \emph{Test Dependencies} section for the github/linguist, which displays the test dependencies for Rust (Cargo), Ruby (Gem), Java (Maven), and Python (PyPI).\footnote{\url{https://andrehora.github.io/testminer/\#github/linguist}}


\subsubsection{Owner and Topic Views}
\begin{figure}[t]
    \centering
    \includegraphics[width=0.43\textwidth]{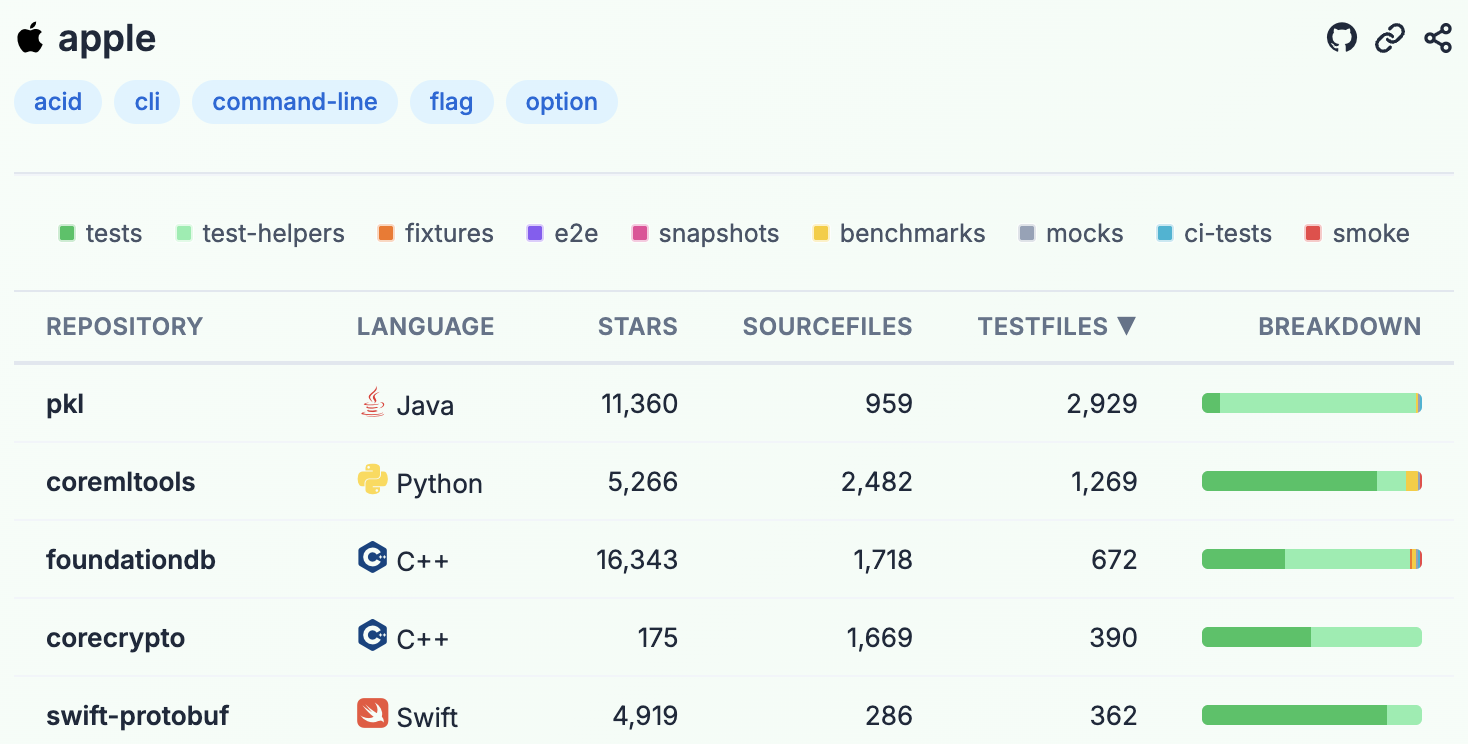}
    \caption{Owner page for Apple.}
    \label{fig:owner-page}
\end{figure}

\begin{figure}[t]
    \centering
    \includegraphics[width=0.43\textwidth]{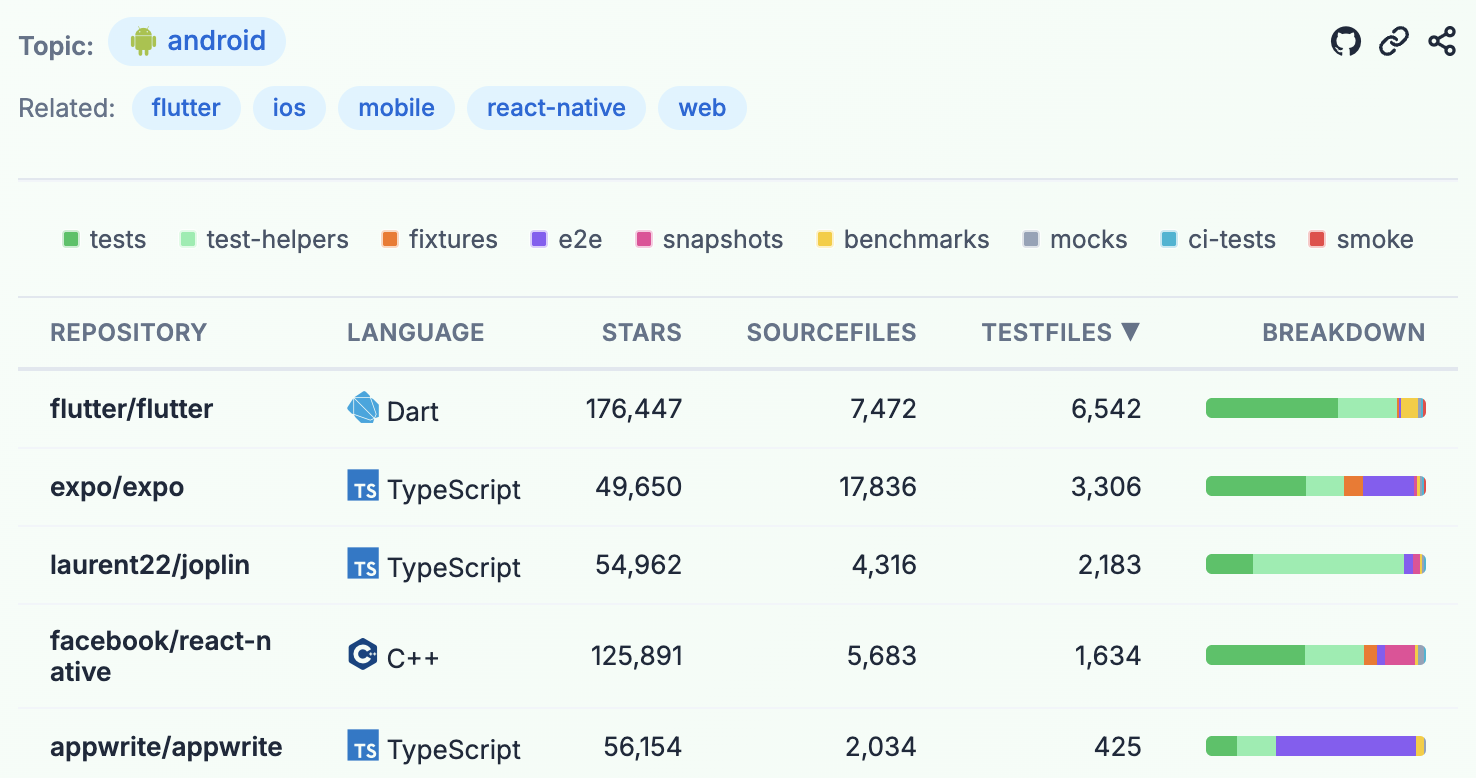}
    \caption{Topic page for Android.}
    \label{fig:topic-page}
\end{figure}

While the Repository View focuses on the analysis of a single repository, the Owner and Topic views enable the analysis of multiple repositories.
Users can specify a GitHub owner (organization or user) or topic.
TestMiner then provides an overview of multiple repositories in a table, enabling users to easily compare them and identify a target repository for further analysis.
Users can also filter results by test category.
Figure~\ref{fig:owner-page} presents the Owner page for Apple\footnote{\url{https://andrehora.github.io/testminer/\#apple}} and Figure~\ref{fig:topic-page} shows the Topic page for Android.\footnote{\url{https://andrehora.github.io/testminer/\#topic:android}}

\subsubsection{Usability and Navigation Features}
Due to space constraints, we can only briefly describe further usability and navigation features of TestMiner:
\begin{enumerate*}[label=(\roman*),font=\itshape]
    \item The \emph{Test Overview} provides complementary data showing test categories grouped by terms, allowing users to easily select a desired test;
    \item Test categories can be mapped to their corresponding source code on GitHub, so by clicking a link in the \emph{Test Overview} or \emph{Test Location}, users can navigate to the corresponding source code on GitHub to inspect the actual test implementation;\footnote{e.g., \url{https://andrehora.github.io/testminer/\#prisma/prisma:cat:tests}}\item In the \emph{Repository View}, users can navigate to a specific release using the dropdown menu;\footnote{e.g., \url{https://andrehora.github.io/testminer/\#prisma/prisma@7.8.0}}
    \item When navigating in TestMiner, the URL is updated, allowing users to share the exact search state;\footnote{e.g., \url{https://andrehora.github.io/testminer/\#prisma/prisma@2.0.1:cat:tests}}
    \item Users can provide a read-only GitHub token to increase the underlying GitHub API limit to 5,000 requests/hour instead of the baseline 60 requests/hour.
\end{enumerate*}
\subsection{Implementation Notes}

TestMiner is a Single Page Application (SPA) built with React for the frontend and Vite for the build process.
The source code is organized into React components, React hooks, and vanilla JavaScript functions. Test data is fetched from GitHub and jsDelivr APIs. 
TestMiner also loads other data, such as repositories for autocomplete suggestions and lists of testing libraries used in SBOM analysis, from local sources.
In this case, the list of repositories is obtained from the SEART GitHub Search Engine (seart-ghs)~\cite{Dabic:msr2021data}.
Moreover, to avoid frequent requests to the GitHub and JSDelivr APIs, TestMiner caches selected testing statistics for the Owner and Topic pages in \texttt{localStorage}.
Finally, TestMiner includes hundreds of unit, component, and end-to-end tests written with Vitest, React Testing Library, and Cypress.


\section{Evaluation}

\subsection{Overview}

We envision TestMiner as a resource for developers, researchers, or contributors seeking to understand a software project's approach to testing.
Computer Science students exploring real-world projects represent a particularly relevant user group and a natural validation context. 
In this context, we conducted a classroom evaluation of TestMiner's usefulness. 
We tasked 50 undergraduate students in a Software Testing course to
\begin{enumerate*}[label=(\arabic*)]
    \item select a real-world GitHub repository,
    \item explore testing practices in the selected project with TestMiner, and
    \item explain a testing practice in an open-ended response.
\end{enumerate*}

\subsection{Qualitative Analysis}

The data collected is publicly available on GitHub~\cite{exploring-software-testing} (each \emph{fork} in the repository corresponds to a student's response). Table~\ref{fig:overview} summarizes the testing practices discussed by participants across the 50 responses (the practices were extracted via thematic analysis).
Next, we describe each category and provide representative examples.

\begin{table}[t]
    \centering
    
    \caption{Summary of testing practices.}
    \begin{tabular}{lrl}
        \toprule
        \textbf{Category} & \textbf{Count} & \textbf{Description} \\ \midrule
        Test organization &	10 & Test organization, location, or structure \\
        Test evolution    & 9 & Test changes over time \\
        Test content    & 6 & Inner workings of tests \\
        Fixtures         & 5 & Test fixtures (e.g, setups, teardowns) \\
        Mocking         & 5 & Test double practices (e.g., mock, fake) \\
        Test type       & 5 & Unit, integration, or e2e tests \\
        Doc test        & 4 & Documentation testing \\
        Edge cases      & 3 & Edge and exceptional cases \\
        Test dependencies &	3 & Dependencies to test libs and frameworks \\
        \bottomrule
    \end{tabular}
    \label{tab:overview}
\end{table}

\subsubsection{Test organization}
This category appeared in 20\% of the responses (10 out of 50).
In this case, students focused on the organization, location, or structure of the analyzed tests.
For example, a student who analyzed \href{https://andrehora.github.io/testminer/#streamlit/streamlit}{streamlit/streamlit} mentioned the overall organization of tests: ``\emph{I chose the Streamlit project because I use it a lot in my daily life and was curious to see how such a popular project organizes its tests. In the TEST LOCATION visualization, it is clear that the highest concentration of tests is in the e2e/ and lib/ folders, and there is also a significant amount in frontend/. This makes a lot of sense, as Streamlit has both a Python part and a heavy web interface}''.
Another student detailed how project \href{https://andrehora.github.io/testminer/#facebook/lexical}{facebook/lexical} structure regression tests when creating bug-reproducing tests~\cite{hora2026understanding}: ``\emph{These tests typically follow the format .../**tests**/\-regression/\-<issue\_\-number>-<issue\_\-subject>.\-mjs, as the file packages/\-lexical-\-play\-ground/\-**tests**/\-regression/\-4876-\-un\-merge-\-cell.\-spec.mjs. Creating test cases for every bug fix is important to prevent regressions and ensure the fix is effective, as the test should fail before the fix and pass after the commit}''.

\subsubsection{Test evolution}

Test evolution occurred in 18\% of the responses (9/50) and refers to focus on how tests evolve and change.
A student described how project \href{https://andrehora.github.io/testminer/#jamesnk/newtonsoft.json}{JamesNK/Newtonsoft.Json} increased its number of test files from 7 to 200 over time: ``\emph{The most interesting data observed in TestMiner for the Newtonsoft.Json project is the constant and linear growth in the number of test files throughout the project's history, from version 1.3.1 to 13.0.4. The project went from 7 test files in v1.3.1 to 200 files in v13.0.4, without any sharp jumps or drops in any version}''.
Similarly, another student observed test growth in \href{https://andrehora.github.io/testminer/#psf/requests}{psf/requests}, noting an increase from 1 to 10 test files: ``\emph{The project started with only 1 test file and reached 10 in the current version. The CI Test (run) only appears from version 2.33.1 onwards}''.

\subsubsection{Test content}

In 12\% of cases (6/50), the responses focused on test content, i.e., the inner workings of tests.
A student analyzed project \href{https://andrehora.github.io/testminer/#nvbn/thefuck}{nvbn/thefuck} (a tool that suggests fixes for incorrect terminal commands) and noted that its tests emphasize Git commands, which are often confusing for users: ``\emph{When observing the test statistics in TestMiner, it was possible to see that a large part of the repository's test files served exclusively to test command fixes for the Git version control system. This focus on Git occurs because, besides being a widely used system, its commands have specificities that can be frustrating for many developers. Thefuck took the opportunity to ensure that most Git commands were automatically corrected and extremely well tested}''.

\subsubsection{Fixtures}

In 10\% of the responses (5/50), students discussed test fixtures, such as \texttt{setUp} and \texttt{tearDown} methods commonly found in testing frameworks.
A student highlighted the number of test fixtures in \href{https://andrehora.github.io/testminer/#fastapi/fastapi}{fastapi/fastapi} and how they contribute to test quality: ``\emph{What caught my attention most was the use of fixtures, with a total of 53 in the project. Fixtures are used to reuse data and configurations in tests, avoiding code repetition. Instead of manually configuring each test, the project defines reusable structures that can be shared among several tests. I found this practice interesting because it makes the tests more organized, easier to maintain, and less repetitive. Additionally, it facilitates writing new tests since many configurations are already ready for use. This contributes to improving software quality by encouraging the efficient creation of more tests}''.
Another student noted that thousands of fixture files were introduced in a specific version of \href{https://andrehora.github.io/testminer/#facebook/react}{facebook/react}: ``\emph{With TestMiner, I noticed that from version 16 to 19, there was a significant increase in the number of tests and especially in the number of fixtures; while version 16 had no fixtures, version 19 has more than 3,500}''.

\subsubsection{Mocking}

In 10\% of the responses (5/50), students explored test double practices, where real components are replaced with simulated versions.
A student mentioned how \href{https://andrehora.github.io/testminer/#soxoj/maigret}{soxoj/maigret} mocks external requests in its tests.
``\emph{Since it makes thousands of external requests, running real tests on all sites would be slow and unstable. The presence of pytest-httpserver and mock indicates that the repository uses Test Doubles. Therefore, the tests simulate site responses to verify that the application can interpret them correctly without needing actual internet access}''.
For project \href{https://andrehora.github.io/testminer/#apache/dubbo}{apache/dubbo}, a student mentioned:
``\emph{The 69 mocks mainly cover simulations of service communication channels and service registries. This makes sense for a distributed communication framework, where testing network behavior directly would be slow, fragile, and difficult to reproduce}''.
Another student mentioned that there are tests for mocks themselves in scikit-learn/scikit-learn: ``\emph{In particular, the use of mocks with the objective of mocking classifiers caught my attention. There is also a battery of tests that verify the functioning of this mock itself}''.

\subsubsection{Test type}

In 10\% of the responses (5/50), students discussed the test types, such as unit, integration, and end-to-end (e2e) tests.
For project \href{https://andrehora.github.io/testminer/#nestjs/nest}{nestjs/nest}, a student highlighted the large number of e2e tests: ``\emph{Something I found very interesting and even counter-intuitive was the large number of e2e tests, even compared to unit tests. After doing some research, I concluded that NestJS has many integration tests because its architecture is based on modules and dependency injection, where correct behavior depends on the interaction between several parts (controllers, services, guards, pipes). Unit tests isolate these parts but do not guarantee that they work well together. Therefore, there is a strong focus on integration tests: they simulate an environment closer to reality and validate the complete flow of the application}''.
A student also mentioned the large number of e2e tests in \href{https://andrehora.github.io/testminer/#stenciljs/core}{stenciljs/core}: ``\emph{The robust multi-platform and end-to-end (E2E) testing strategy stands out. The presence of 32 E2E test suites, combined with the use of tools like Puppeteer and Playwright (indicated by the dependencies), demonstrates that the team prioritizes validating component behavior in real browsers}''.

\subsubsection{Other cases}

Documentation testing, edge cases and test dependencies were mentioned less frequently.
A student who analyzed \href{https://andrehora.github.io/testminer/#fastapi/fastapi}{fastapi/fastapi} mentioned the use of tests as a means of documentation, mitigating the risk of outdated documentation: ``\emph{What caught my attention most was the density of the tests/test\_tutorial directory, which leads with 221 files. The fact that the highest concentration of tests is in the test\_tutorial folder reveals that the code snippets and examples in the official documentation and tutorials are actually the framework's own automated tests. This significantly reduces the risk of outdated documentation, as any update that breaks a feature will cause the tutorial example to fail. It is an interesting practice that makes perfect sense}''.




\begin{boxH}
\noindent\textbf{Summary:}
Students critically explored a diverse range of real-world testing practices, drawing conclusions that would have required substantial manual effort without tool support. Qualitative responses suggest that TestMiner lowers the barrier to engaging with unfamiliar codebases, supporting its potential as both a practitioner and educational tool.
\end{boxH}

\section{Limitations}

TestMiner relies on the GitHub API to collect repository metadata and is therefore subject to a limit of 60 requests/hour. If users reach this limit, some features may become partially unavailable. TestMiner mitigates this limitation by accepting a read-only GitHub token, which increases the rate limit to 5,000 requests/hour; this token is stored in the browser’s {\small\texttt{localStorage}}.
Second, GitHub links in the \emph{Test Overview} and \emph{Test Location} sections may occasionally not work due to synchronization inconsistencies between the data returned by the API and the actual codebase on GitHub, particularly in the \texttt{main} and \texttt{master} branches.
Third, the API may return partial information for very large repositories, which can lead to incomplete test statistics.
In such cases, TestMiner displays a message informing users that this is an API limitation.

\section{Related Work}

The literature provides multiple tools for analyzing Git and GitHub repositories, e.g.,~\cite{spadini2018pydriller, gitevo, Dabic:msr2021data, gitpython, isomorphic-git, jgit}.
Commonly, these tools allow users to mine repository metadata and source code, typically via command-line and live datasets, but without specific emphasis on testing practices.
Commercial tools such as CodeCov~\cite{CodeCov} and SonarQube~\cite{SonarQube} are powerful, but require CI instrumentation.
TSDetect~\cite{tsDetect} and PyNose~\cite{wang2021pynose} surface insights about test quality, but their use is confined to detecting ``smells'' in unit tests.
Unlike these tools, TestMiner is a web application focused on the analysis of software testing, allowing developers to navigate and explore GitHub repositories through a testing-oriented perspective.

\section{Conclusion}

We presented TestMiner, a tool for exploring software testing practice on GitHub projects.
As future work, we plan to incorporate source code analysis into TestMiner, potentially leveraging {\small\texttt{tree-sitter}} parsers \cite{tree-sitter} to support multiple programming languages.
We would also like to explore the use of Large Language Models (LLMs) to make the output of TestMiner even more accessible to users, e.g., by generating a concise natural language report highlighting the key testing practices observed in the user's provided repository.

\section*{Acknowledgments}

Research supported by CNPq, CAPES, and FAPEMIG.

\bibliographystyle{IEEEtran}
\bibliography{main}

\end{document}